\documentclass[12pt,a4paper]{article}
\pdfoutput=1

\usepackage[utf8]{inputenc}
\usepackage{textcomp,marvosym}
\usepackage{amsmath,amssymb}
\usepackage{graphicx}
\usepackage{natbib}
\usepackage{booktabs}
\usepackage[affil-it]{authblk}
\usepackage{breakurl}
\usepackage[obeyspaces]{url}
\usepackage[breaklinks]{hyperref}

\usepackage[labelfont=bf,labelsep=period,justification=RaggedRight,%
            singlelinecheck=off,textfont=small]{caption}

\begin{document}

\title{Catastrophic Regime Shift in Water Reservoirs and S\~ao Paulo Water Supply Crisis}
\author[1]{Renato M.  Coutinho
    \thanks{Electronic address: \texttt{renatomc@ift.unesp.br}; Corresponding
author}
}
\author[2]{Paulo I.  Prado
    \thanks{Electronic address: \texttt{prado@ib.usp.br}}
}
\author[1]{Roberto A. Kraenkel
    \thanks{Electronic address: \texttt{kraenkel@ift.unesp.br}}
}

\affil[1]{Instituto de F\'{\i}sica Te\'orica,  Universidade Estadual Paulista -
        UNESP, S\~ao Paulo, Brazil}
\affil[2]{LAGE do Departamento de Ecologia, Instituto de Bioci\^encias,
Universidade de S\~ao Paulo - USP, S\~ao Paulo, Brazil}

\maketitle
\renewcommand{\thefootnote}{\fnsymbol{footnote}}
\footnotetext[0]{The authors contributed equally to this work.}

\begin{abstract} The relation between rainfall and water accumulated in
    reservoirs comprises nonlinear feedbacks.  Here we show that they may
    generate alternative equilibrium regimes, one of high water-volume, the
    other of low water-volume.  Reservoirs can be seen as socio-environmental
    systems at risk of regime shifts, characteristic of tipping point
    transitions.  We analyze data from stored water, rainfall, and water
    inflow and outflow in the main reservoir serving the metropolitan area of
    S\~ao Paulo, Brazil, by means of indicators of critical regime shifts, and
    find a strong signal of a transition.  We furthermore build a mathematical
    model that gives a mechanistic view of the dynamics and demonstrates that
    alternative stable states are an expected property of water reservoirs. We
    also build a stochastic version of this model that fits well to the data.
    These results highlight the broader aspect that reservoir management must
    account for their intrinsic bistability, and should benefit from dynamical
    systems theory. Our case study illustrates the catastrophic consequences
    of failing to do so.
\end{abstract}

\section{Introduction}

Complex socio-ecological and socio-environmental systems often involve interactions between natural elements and human action in a nonlinear and adaptive way \citep{folke2006}.  Such systems have received much attention \citep{gordon2008,osterblom2013} recently and may display many of the common dynamical features present in natural systems.  Here we will be interested in the existence of alternative stable states, representing different possible dynamical regimes, and transitions between them.  These tipping-point transitions have been well studied and  are common in natural systems \citep{scheffer2001}. Desertification \citep{klausmeier1999, meron2004, scheffer2003} and lake eutrophication \citep{scheffer2007, carpenter2005} are only the most notable ones among a plethora of cases.  Characterization of a transition from time-series data can be assessed using indexes related to either return-time to equilibria or variability near the tipping point \citep{dakos2012}.  These techniques, together with models that encompass the main features of the system, may be extended to systems where the  interaction with the human factor  creates the possibility of a certain level of control or, at least, of action attempting to bring the system to a situation that is considered desirable.  Our attention will be focused on the dynamics of water reservoirs, which has not been studied under this light, and on the characterization of alternative stable states and associated transitions. We show a case in which these dynamic properties had been neglected in the management of one of the largest water reservoirs in the world, the Cantareira system, which serves the Metropolitan Area of S\~ao Paulo (MASP). 

\begin{figure}[htb]
\centering
    \includegraphics[width=0.9\textwidth]{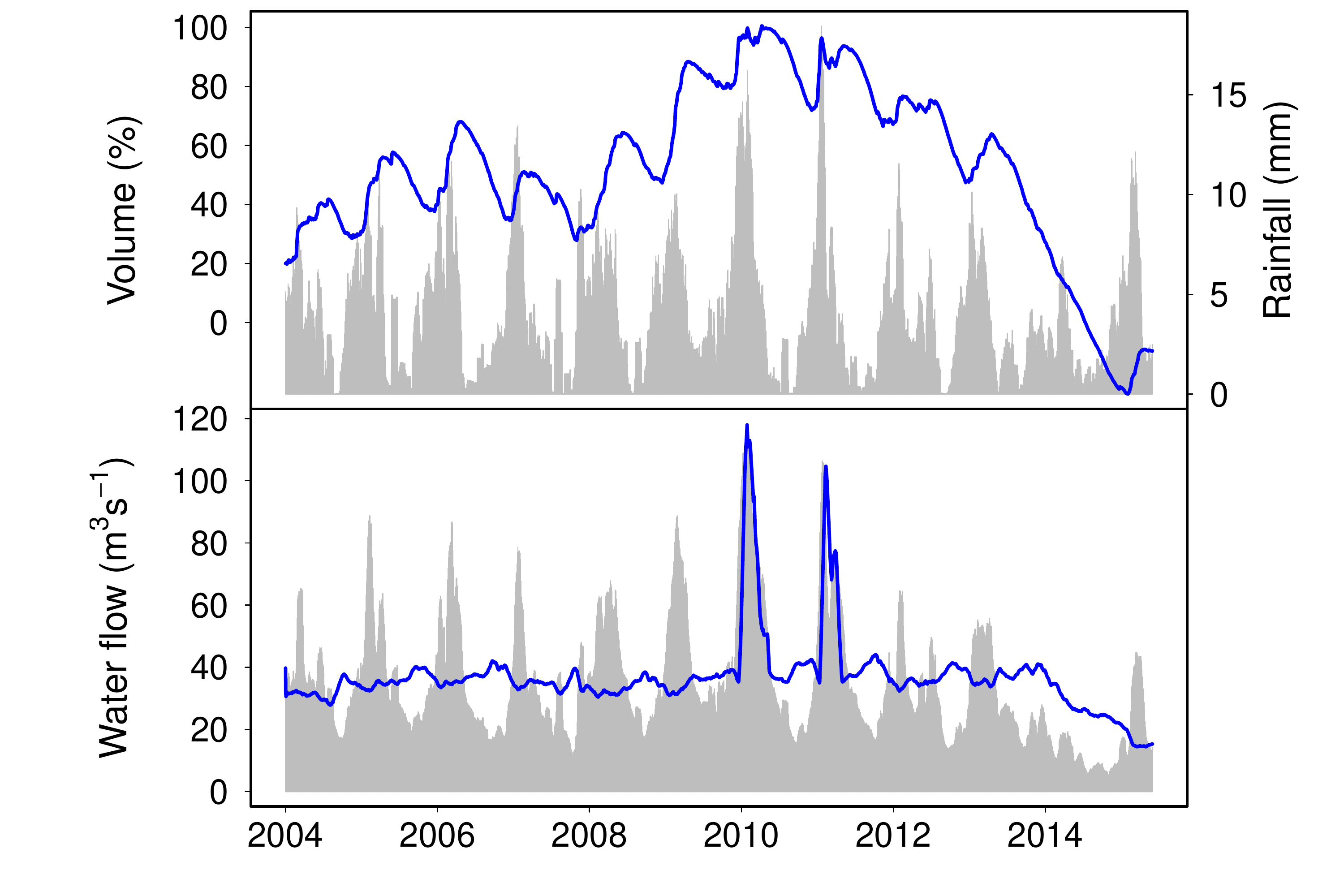}
\caption{{\bf Rainfall, water flows and volume stored in the Cantareira system
since 2004.}  \textbf{Upper panel:} notwithstanding the inter-annual trend, a
clear seasonal fluctuation is present in the rainfall (gray) which was
followed by the volume of stored water (percentage of operational volume, blue
line) until mid 2013. From this point on, the volume of water decreased to
very low levels, without any seasonal trend.  A recovery of small amplitude is
seen from Feb-May 2015 due to reduction in outflow. 
This is shown in the \textbf{lower panel:} the water inflow into the reservoir (gray) showed seasonal fluctuations 
coupled to rainfall. The reservoir operators kept outflow (blue line) around the maximum 
average allowed (36 $m^3s^{-1}$, exceeded only to avoid overflow). 
Operators started a gradual decrease in outflow in the beginning of 2014, when the operational volume was close to depletion. 
Since the depletion (July 2014), water is available only by pumping, 
and outflow reduction proceeded in a slow pace till January 2015, when it was abruptly cut by 25\%.}
\label{volume}
\end{figure}

MASP faces an unprecedented crisis of water supply. This exceptional situation can be readily seen from the plot of the volume of water stored the Cantareira system, the main reservoir serving the area, over the past twelve years (Fig.(\ref{volume})).  The volume of water decreased sharply from mid-2013 and the operational capacity of the reservoir was depleted in July 2014. Since then water withdrawal is done by pumping of the so-called ``strategic reserve'' or ``dead volume''. The S\~ao Paulo Water Company, SABESP, began to reduce withdrawals in January 2014, and by May 2015 the total outflow was 40\% of the average values. Additionally, the last rainy season (October 2014 -- March 2015) provided more rainfall compared to the previous two years. Despite that, only of 15\% of total volume had been recovered, and the reservoir remains operationally exhausted.  This situation seems typical of catastrophic regime shifts, driven by a tipping-point transition.  In this work, we will establish this regime transition on firm grounds and discuss its consequences.  For sake of brevity, we will use the term {\it reservoir} to mean a system of interconnected reservoirs.

We propose that reservoir water-volume is subject to drastic regime changes due to the underlying bistability of the system. Bistability is the condition in which there are at least two possible stationary states of a system for the same set of parameters. In our case, the parameter is rainfall and the variable that may assume two possible alternative stable states is the volume of water stored in the reservoir. If the reservoir is at high levels the catchment has more stored water and thus more of the rainfall will flow into water the reservoir. On the other hand, for low water levels the most obvious change is that much of the water is absorbed by dry soil and thus a lower proportion of rainfall becomes stored water in the reservoir. Many other factors may be play and the exact relation between rainfall and the amount of water stored in the reservoir is an instance of the rainfall-runoff problem in hydrology \citep{beven2011}.  What matters for our discussion is that these feedback mechanisms are drivers of critical transitions.

Both states, high-level and low-level regime, are resilient in ranges of rainfall and outflow parameters that may overlap, leading to hysteretic behavior: the paths to and from the low-level state are not the same. Resilience of the high-level water volume regime may be lost, for instance, if the rainfall index falls below a certain critical value, or if the reservoir is over-exploited. The system then transits to the low-level regime, in a tipping-point transition entailing a regime shift. The dramatic side of this fact is that the low-level regime is also resilient, so either rainfall index has to increase  beyond the mentioned critical value, or outflow has to be reduced drastically in order to produce a backwards transition to the high-level regime.  This state of affairs prompts us to characterize transitions by indexes related to the critical slowing down phenomenon \citep{dakos2015}. These indexes are usually connected to research of early-warning signals of critical transitions. Discussions in recent literature have raised doubts about how early the signals are \citep{boettiger2013}, but this is not actually relevant in the present case, as we are concerned mainly with characterization and not anticipation of the transition.

In what follows we proceed by the following steps: we first inspect the data to establish the existence of a feedback mechanism; next we look for indicators of critical transitions in two steps, first by a qualitative assessment of the data, and second by demonstrating the existence of a peak in the variance characteristic of near tipping point regions.  We then turn to a further step by proposing a mathematical model for the dynamics of the system, in two versions. First, a deterministic version where we can understand the dynamical behavior of reservoirs.  Next, we propose an stochastic extension of the model, via a stochastic differential equation which can be fitted to actual data accurately and used for short-term extrapolations.

\section{Model Independent Results}

\subsection*{The feedback mechanism}
\begin{figure}[htb]
\centering
    \includegraphics[width=0.82\textwidth]{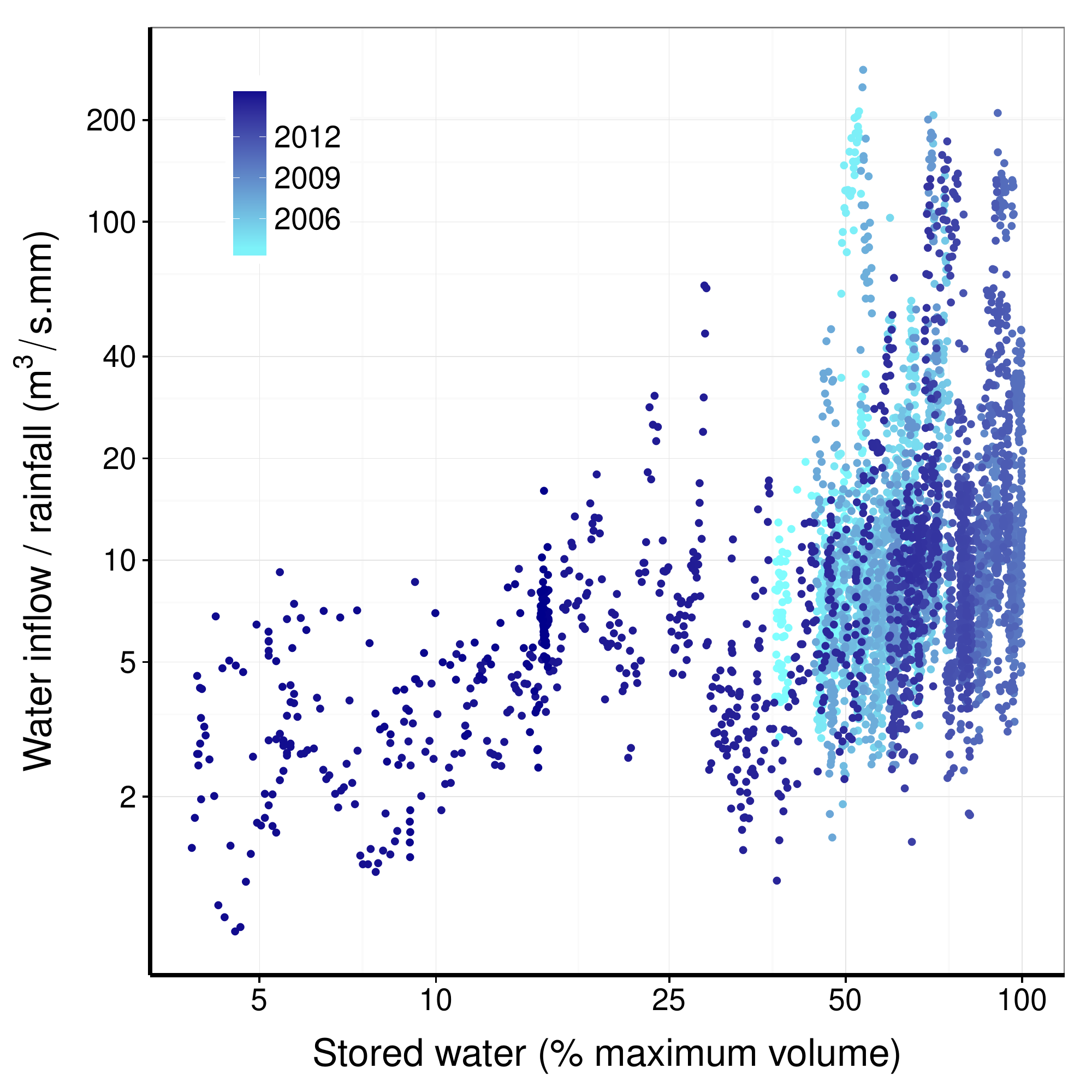}
\caption{{\bf Dependence of the ratio inflow to rainfall on the stored water in the reservoir.} The Cantareira reservoir shows the well-known non-linear transfer of rainfall to runoff. The volume of water in the reservoir is a surrogate of total water in the whole catchment. The higher the volume the wetter the catchment and the larger the inflow/rainfall ratio. The data is the same of Fig. \ref{volume}.}
\label{mech}
\end{figure}

Given a certain amount of rainfall, how much water will effectively flow into the reservoir? This is connected to the rainfall-runoff problem, central in hydrology.  The big picture is that the wetter the catchment the more of the rainfall is converted in runoff water \citep{beven2011},  that in this case will flow to the reservoir. A complete assessment of this situation involves modeling the rainfall-runoff relation in an explicit spatial setting, considering soil heterogeneities, terrain and many other variables. However, in order to proceed in a simpler way, we postulate that the volume of water stored in the reservoir is a surrogate for the drainage basin condition.  To verify that this is a sound assumption, we show in Fig.(\ref{mech}) the ratio of water inflow to the rainfall in the Cantareira reservoir as a function of the volume, yielding a clear trend: the higher the volume, the more efficiently the rainfall becomes water flowing into the reservoir.

\subsection*{Assessment of a critical transition}

\begin{figure}[htb]
\centering
    \includegraphics[width=0.9\textwidth]{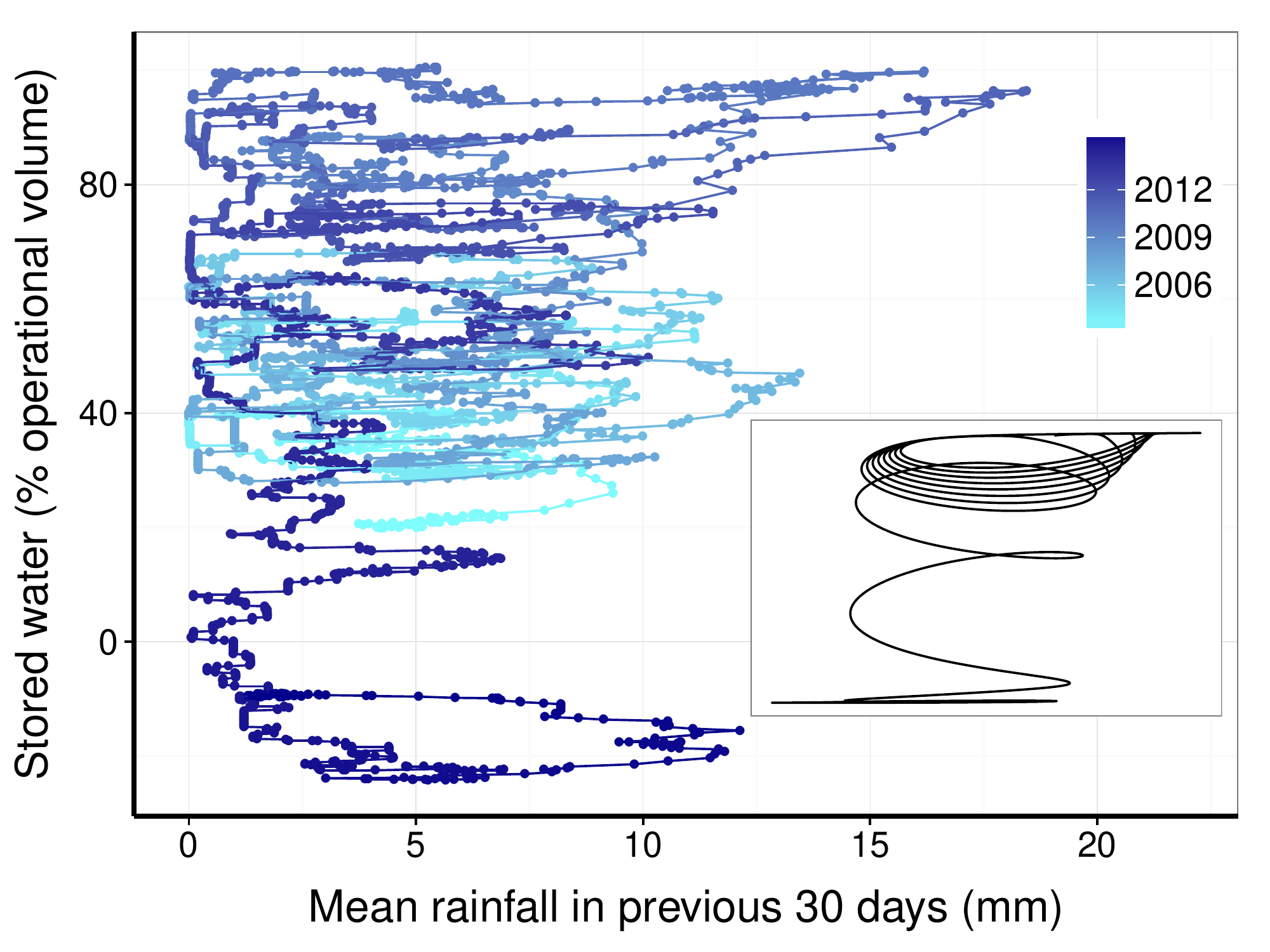}
\caption{{\bf Observed and simulated transition between two dynamical regimes in the rainfall-volume plane.} In the main panel we show the actual data for the Cantareira system. In the inset we show the results of a differential equation describing this system, which exhibits bistable behavior. The equation was simulated with variation of rainfall with a time oscillating part plus a long-term decreasing trend.}
\label{funil}
\end{figure}

In Fig.(\ref{funil}) we show the data for the Cantareira reservoir in the rainfall \emph{vs.} stored-water plane. In normal situations, we would expect to have nearly closed curves in the upper part of the plane. These curves come from the seasonality  of rainfall, which induces  similar oscillations in the water stored in the reservoir.  However, what is seen in the plot is that from 2013 on, the system spins down to a new cycle on the lower part of the plane. This lower cycle represents, again, seasonal variations, but the volume now oscillates around much smaller values.  In the inset of Fig.(\ref{funil}) we show what would be the result obtained from an ordinary differential equation displaying bistable behavior, in which the rainfall parameter oscillates in time and has a decreasing trend, inducing a regime change. Similarities are evident. The specific equations that generate Fig.(\ref{funil}) will be discussed in the Mathematical Models section.

\begin{figure}[htb]
\centering
    \includegraphics[width=0.9\textwidth]{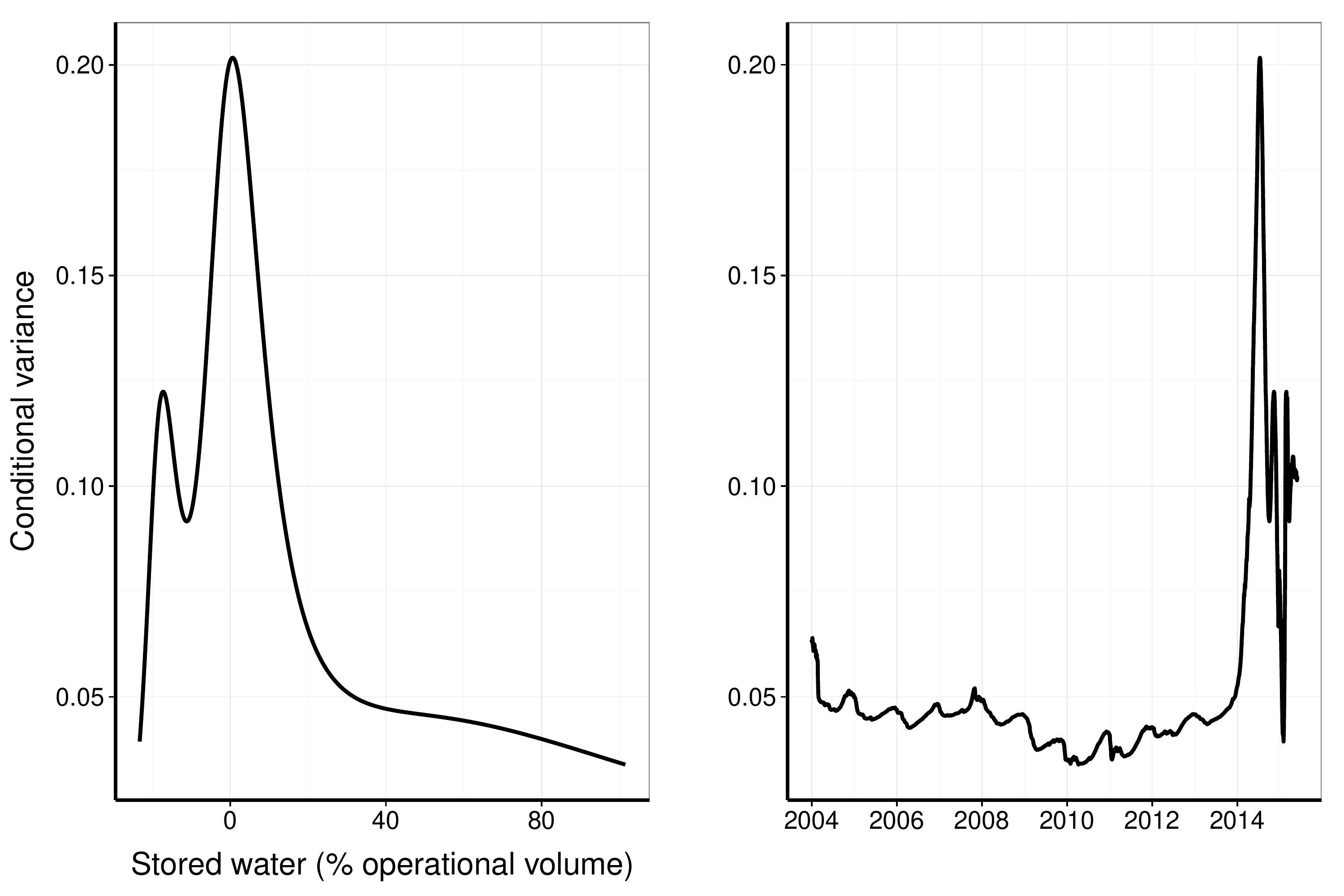}
\caption{{\bf Assessment of critical regime shift in a water reservoir by a peak in the conditional variance}. 
Conditional variance from a drift-diffusion-jump model (\emph{ddj}) for the
Cantareira water volume as a function of relative stored volume (\textbf{left}), and of time (\textbf{right}).
The \emph{ddj} model was fitted to the daily time series of the 
relative stored volume, log-transformed as recommended by \citet{dakos2012}.
The time series (\textbf{right panel}) shows an abrupt increase of \emph{ddj} conditional variance 
at the beginning of 2014, which indicates that a state transition occurred. The \textbf{left panel} shows that the 
high variance values occurred in a narrow range of low volumes in the reservoir, 
recorded in the end of 2013 and beginning of 2014.}
\label{cond}
\end{figure}

The previous discussion gives a qualitative view of a transition, but we would like to take a step forward and provide a more quantitative assessment. In recent years, a toolbox for detecting early-warning signals of critical transitions has been developed \citep{dakos2012,dakosews}. These are based on the critical slowing down phenomenon, implying that the lag-auto-correlation function and variance indicators should increase near the transition.  As discussed in \citet{boettiger2012}, a series of underpinnings exist concerning the reliability of these indicators, which may fail to anticipate the occurrence of a transition. However, in this work we are interested in characterizing the regime shift rather than anticipating it. In order to do so, we evaluated evidence of 
a critical transition with a diffusion-drift-jump model (\emph{ddj}).
If the time series is long enough and has high frequency of observations, this model gives
an accurate description of the underlying process that generates the data \citep{carpenter2011}.
Variations along time are described as a combination of instantaneous changes by deterministic trends (the drift), 
instantaneous variation (diffusion) and occasional uncorrelated changes (jumps). 
The model also allows to estimate the variance conditioned to these components. 
In theory, this conditional variance goes to
infinity at bifurcation points of the underlying dynamics \citep{dakos2012}. 
We found a very clear peak in the conditional variance near the beginning of 2014 (Fig.(\ref{cond})).
This gives the first characterization of an indicator of critical regime shifts for water reservoirs.
Running-window statistics like variance and autocorrelation at lag one \citep{dakos2012} followed the same trend.

As discussed in \citet{dakos2015}, critical transitions are better assessed when, 
together with a statistical indicator, like the conditional variance used in Fig.(\ref{cond}), a  mechanism leading to 
bistability is evident. The feedback discussed above is the basic mechanism we hypothesize.  A further step, therefore, is to build a mathematical model that takes into account the feedback and may be fitted to the data, and show that it exhibits bistability. This is what we present in the next section.

\section{Mathematical models for reservoir dynamics}

\subsection*{Deterministic model}
The change in time of stored-water volume in a reservoir is determined by inflow and outflow rates.  Outflow rates are a consequence of management decisions. Inflow comes ultimately from rainfall through a set of hydrological processes.  This leads to an equation of the form:

\begin{equation}
\frac{dV}{dt} = - s(V) + r(V,R) ~,
\end{equation}
where $V$ is the volume of water stored in the reservoir, $s(V)$ is the outflow rate and $r(V,R)$ is the inflow rate, with $R$ a measure of the rainfall. We now discuss these terms separately.

\begin{figure}[htb]
\centering
    \includegraphics[width=0.85\textwidth]{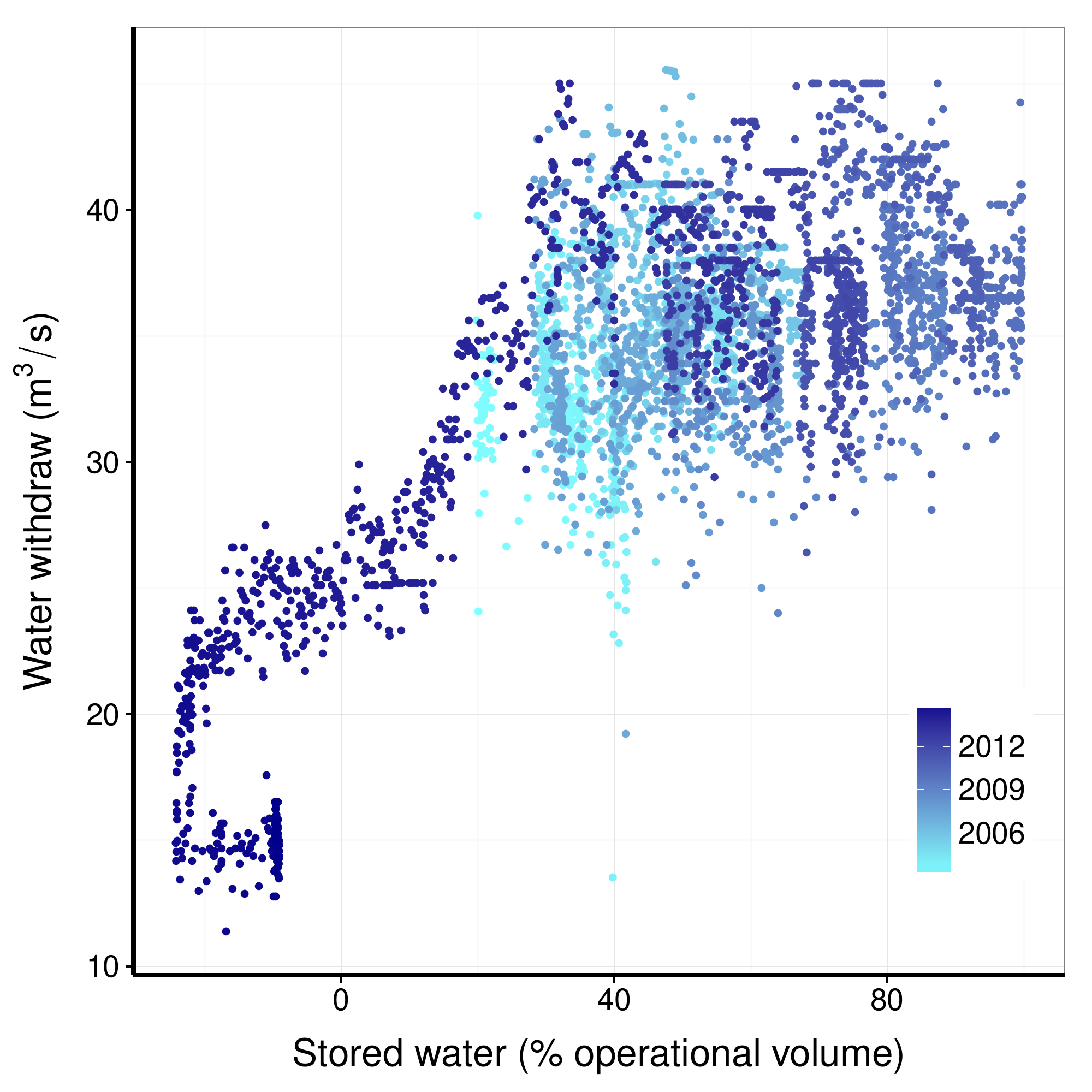}
\caption{{\bf Water withdrawal from the Cantareira reservoir depends on stored water volume.} Withdrawals by the reservoir operators  saturated to a maximum average value when stored water was above one third of operational capacity. Note also that when stored water fell below -15\% operators plunged withdrawals from ca. 23 to 14~$m^3s^{-1}$.
Since then releases are kept at these levels despite an increase in the stored volume, 
to save the current volume (darker points). } 
\label{out}
\end{figure}

\paragraph{Outflow.} We assume that withdrawal of water is managed and saturates to a maximum, $s_0$,  when the stored volume is high, but is reduced as water becomes scarce (which could be done both to prevent complete disaster or due to the difficulty in pumping water from low levels). Such assumption is realistic for the Cantareira system: while stored volume was above 33\% water withdrawals oscillated around 36~$m^3s^{-1}$, according to the maximum quota set by the Brazilian Water Agency, \emph{ANA} (Fig\ref{out}).  As the volume dropped outflow decreased because \emph{ANA} pushed the main reservoir operator SABESP to cut down withdrawals. To model this situation we use the following simple form:
\begin{equation}
 s(V) = \frac{s_0 V}{K+V} ~, 
\end{equation}
where $K$ is the volume of water in which the withdrawal is reduced to $s_0/2$.  It is a measure of how cautious the water management is. 

\paragraph{Inflow.} As already implied in the notation $r(V,R)$, inflow depends not only on rainfall, but also on volume. As discussed above, we propose that inflow is higher for higher volumes.  A simple assumption is thus:

\begin{equation}
r(V,R)= \begin{cases}
        \alpha R^{\beta} V^{\gamma} &\text{, if } V < V_{max} \\
        0 & \text{, otherwise}
  \end{cases}\end{equation}
where $\alpha$, $0<\beta \leq 1$ and $0<\gamma\leq1$ are phenomenological parameters. We also introduced a maximum volume for the reservoir, $V_{max}$. Once it is attained, no increase in stored water is possible and excess inflow is released downstream through a spillway. 

\paragraph{}Putting everything together, we come to the following differential equation:
\begin{equation}
\frac{dV}{dt}= 
        -  \frac{s_0 V}{K+V}  + \alpha R^{\beta} V^{\gamma} ~,
\label{determ}
\end{equation}
if $V<V_{max}$. If $V>V_{max}$, the second term on the right-hand side is zero.
Equation (\ref{determ}) is not meant to be a full model of all hydrological processes, but rather a description of the basic dynamical factors affecting the volume of water stored in a reservoir.  In spite of its simplicity, a stochastic version of it fits very well to observed data, as will be seen below.

\begin{figure}[htb]
\centering
    \includegraphics[width=0.8\textwidth]{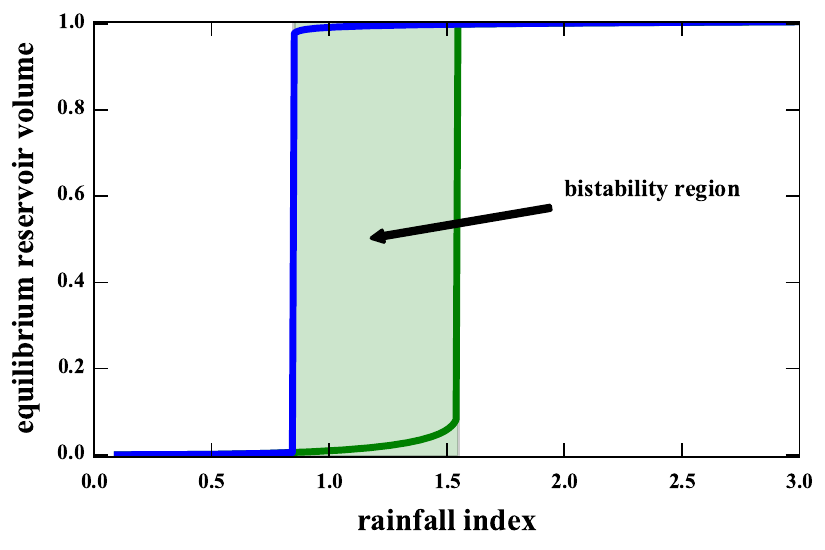}
\caption{{\bf Fixed points of Equation (\ref{determ}) as a function of rainfall $R$.} We assumed that $V_{max} =1$, and $V$ is measured in fractions of the maximum volume. Parameter values are: $\alpha =1$, $s_{0} =1$, $K =0.2$, $\beta = 1$, $\gamma =2/3$. The region of bistability is shadowed.}
\label{bistable}
\end{figure}

If the rainfall is constant over time, Equation (\ref{determ}) has either one or two stable fixed points, depending on the value of $R$.  Fig.(\ref{bistable}) shows the general situation, assuming specific values for the parameters. The existence of a bistability region is clear.  

The existence of the bistability region  depends essentially on the fact that inflow is dependent on the volume. The form of the outflow term is not essential: all that is required is that it goes to zero as $V$ goes to $0$ and that it saturates at some value. The existence of a maximum volume is, from the mathematical viewpoint, a form of avoiding arbitrarily large volumes.

\begin{figure}[htb]
\centering
    \includegraphics[width=0.8\textwidth]{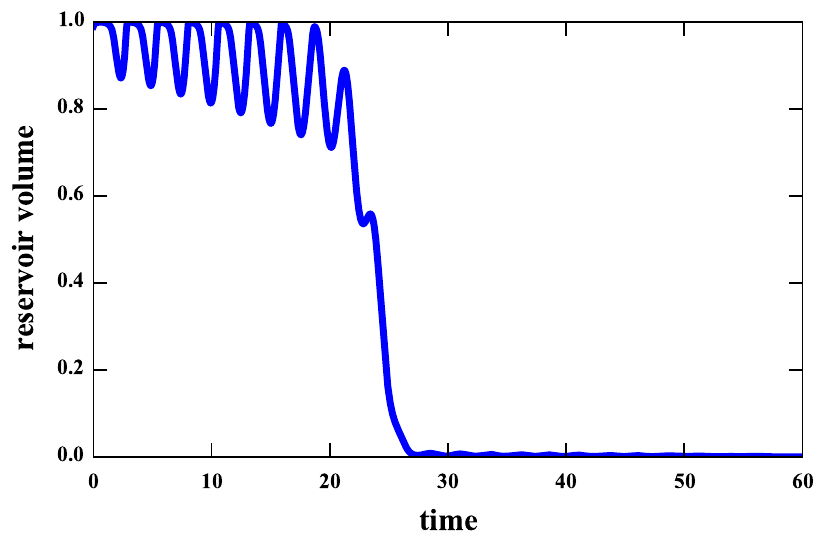}
    \caption{{\bf Regime shift in our mathematical model.} The volume time series is a representative numerical solution of equation \ref{determ}. For this solution rainfall is assumed as $R(t)=R_0(1+a\sin(2\pi ft))(1-bt)$, with $R_0=1.1$,  $a=0.4$,  $f=0.4$ and $b=0.01$. All the other parameters are the same as in Fig.(\ref{bistable}). The parametric plot $V(t)$x$R(t)$ for this solution is given in the inset of Fig.(\ref{funil}).}
\label{solut}
\end{figure}

Obviously, in real situations $R$ is not constant. It has clear seasonal variations and may be subject to long-term trends.  Solutions, in this case, will not tend to a constant value. Bistability will be reflected in the existence of different regimes, corresponding to oscillations around the fixed points of the time-independent problem, as long as the seasonal variations and trends are not too large. In Fig.(\ref{solut}) an example is shown, with $R=R_0(1+a\sin(2\pi ft))(1-bt)$. We also plot the solution as a curve in the $V$x$R$-plane (inset of Fig.(\ref{funil})), showing the distinctive feature of oscillations in the upper part, spinning down on a short time scale to the lower part. This pattern is present also in the observed relationship between rainfall and stored volume in the reservoir (main panel of Fig.(\ref{funil}).

\subsection*{Stochastic model}

In order to compare theory to data we need to take a further step and allow for stochastic fluctuations. This leads us to consider the following stochastic differential equation:

\begin{equation}
  \label{stoch}
  dV =  ( \alpha R^{\beta} \, V^{\gamma} \; - \, E)\, dt \, + \, \sigma V \,dW ~,
\end{equation}
where $E$ is the outflow of water ($m^3s^{-1}$), $R$ is the mean rainfall in the previous 30 days (mm)
 and  $W$ is the standard Wiener process.
The term $\sigma V dW$ expresses an instantaneous stochastic Gaussian noise in the stored volume,
with zero mean and standard deviation proportional to the stored volume.
We also assume that the observed values of stored volume follow a Gaussian distribution with mean value 
equal to the expected value of the stochastic equation 
and an unknown standard deviation $\epsilon$.  The reason to take $R$ as a 30-day mean is empirical: volume does not respond instantaneously to rainfall, and complex hydrological processes tend to spread effects over time. This non-locality in time is encompassed by taking a mean rainfall over a certain period. The period itself (30 days) is  justified {\it a posteriori} as the one giving the best fit to data.

Equation (\ref{stoch}) keeps the deterministic skeleton from the previous discussion, at the same time allowing us to take $V$, $R$, $E$ and inflow from time-series of recorded data for the Cantareira reservoir. We recall that our model predicts catastrophic shifts  because water inflow depends on the stored volume, which
creates a feedback mechanism. To evaluate the support provided by data to this hypothesis, we
evaluated a competing model in which inflow depends only on rainfall:

\begin{equation}
  \label{alternative}
  dV =  ( \alpha R^{\beta} \, - \, E) dt \, + \, \sigma V dW
\end{equation}

The dataset are the time series of daily inflow, outflow, rainfall, and stored water
volume. The model provided good fits for time spans up to one year.
In this paper we focused on the last 365 of the time series (May 31 2014 to May 31 2015). 
Because simultaneous fitting of all five parameters ( $\alpha$, $\beta$, $\gamma$, $\sigma$ and $\epsilon$) did not converge
or provided unreasonable estimates, we callibrated both models in 3 steps:

\begin{enumerate}
\item The observation error $\epsilon$ was conservatively estimated from trajectory matching \citep{pomp}.
\item The exponent $\gamma$ was estimated as the slope of a Gaussian linear regression of reservoir area
  in function of reservoir volume in log scale.
\item Parameters $\alpha$, $\beta$ and $\sigma$ were estimated with Bayesian particle filter \citep{liu2001}
  modified by \citet{pomp}.
\end{enumerate}

\begin{figure}[htb]
\centering
  \centering
  \includegraphics[width=0.9\textwidth]{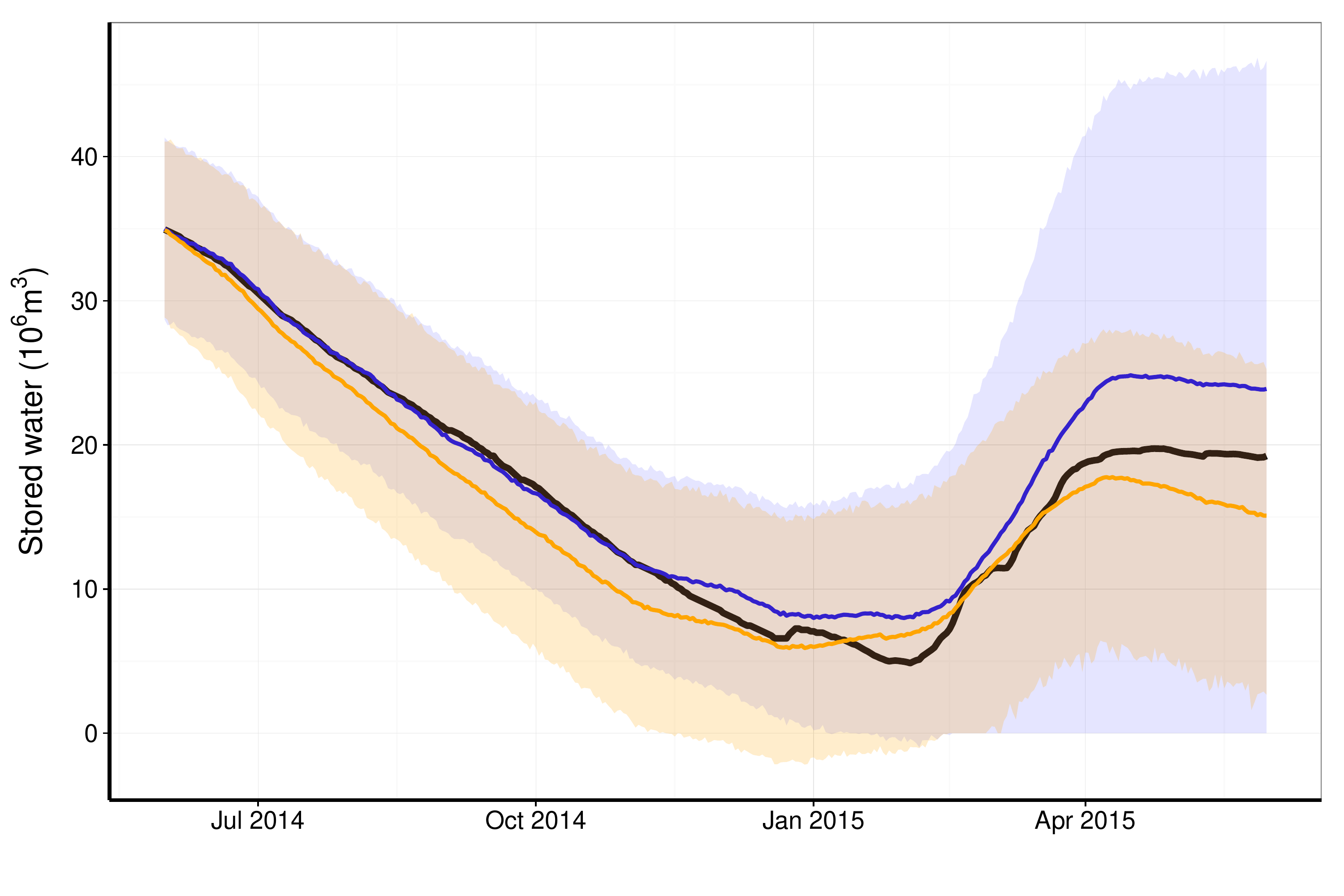}
  \caption{{\bf Time series of observed and predicted stored water volume in the Cantareira reservoir system.} 
  The black line shows the observed time series which are daily records from 2014-05-31 to
  2015-05-31. The orange lines and shaded areas show 
  the expected trajectories and Bayesian credibility interval from  a mathematical model 
  of observed inflow as function of rainfall (Equation \ref{alternative}). The blue lines and shaded areas
  show the expected values and intervals from a model of inflow as function of both rainfall and the stored volume 
  (Equation \ref{stoch}). Trajectories and intervals were calculated from 5000 numerical simulations
  of each model, with parameters sampled from the posterior distribution of the Bayesian particle filter estimation.
}
\label{fit}
\end{figure}

Equation \ref{stoch}, which describes expected water inflow as a function of rainfall and current stored volume,
provided a much more plausible fit to the observed time series (log-likelihood ratio: 12.11, Tab. \ref{tab:fits}). 
The model that does not take into account the effect of volume on inflow
(Eq.(\ref{alternative})) underestimated the stored volume in most of the period, and
did not predict the increase and further stabilization of stored volume since February 2015 (Fig.(\ref{fit})).
The better fit of equation \ref{stoch} supports the hypothesis that the ratio of inflow to rainfall depends on
the volume. This feedback is caused by the interaction between rainfall and stored volume, a surrogate 
of the hydrological state of the catchment. 
This, in turn, substantiates our statement about the existence of alternative states due to a feedback process.

\begin{table}[h!]
  \centering
  \caption{Estimates of parameters and Log-likelihood (LL)
    for two competing models to predict the changes in stored water volume in Cantareira Reservoir.
    The model that describes water inflow as a function of rainfall and stored water 
    ($f(\text{rain},\text{volume})$, Eq.(\ref{stoch})) had a much higher likelihood and thus provides a 
    much more plausible description of the time series.
  }\vskip0.5cm
  \begin{tabular}{lcccccc}
    \toprule
    \textbf{Model} & \textbf{$\alpha$} & \textbf{$\beta$} & \textbf{$\gamma$}& \textbf{$\sigma$} & \textbf{$\epsilon$} & \textbf{LL}\\
    \midrule
    Eq.\ref{stoch}:$f(\text{rain},\text{volume})$ & 5.998 & 1.043  &0.590 &0.00231 & $3.2\times10^7$ & -6676.0\\
      Eq.\ref{alternative}:$f(\text{rain})$& 392494.3 & 0.926  & -- & 0.00948 & $3.2\times10^7$& -6688.1\\
    \bottomrule
  \end{tabular} 
  \label{tab:fits}
\end{table}

\section{Concluding remarks}

We have demonstrated that the dynamics of the volume of water stored in a reservoir is subject to abrupt regime shifts and hysteretic behavior. These results follow from: (i) a qualitative assessment of the dynamics; (ii) a quantitative  evaluation of indexes characterizing regime shifts; (iii) a modeling approach that gives a mechanistic view of the dynamics; and (iv) a fit of a corresponding stochastic model to the data.

These results have serious significance for water reservoirs management.  Resilience of the normal operating conditions may be lost under drought conditions in favor of a new low water levels regime, which is itself resilient under a range of external conditions. Recovery of the normal condition can, thus, be difficult, and may demand potentially extreme measures. These measures are those that allow for decrease in the outflow of the reservoir, representing a possible burden to populations served by the reservoir. This implies that cautious management of a reservoir should avoid the regime shift.  The major question then becomes: how to achieve such management?  We have taken a first step here, identifying the need to acknowledge the potential for such catastrophic transitions.  Under this perspective, early warning signals of catastrophic regime shifts are a possible avenue to tackle the problem in the absence of detailed hydrological data, but the extent of their predictive power has yet to be ascertained.

Studies in water resources have mainly favored full-blown models to depict the ever increasing understanding of hydrological processes \citep{beven2011}. 
Nevertheless simple mathematical descriptions of the first principles that operate in catchments also proved to be useful 
both in fitting data and understanding the underlying mechanisms \citep{beven2011, young2004}. 
We followed this synthetic approach to propose a model akin to those used in system dynamics. 
Our models are a bare-to-the-bones description on how rainfall turns into stored water in a reservoir. In doing that we depict many hydrological processes in a few fixed parameters. What could remain from such a phenomenological approach are very general dynamic properties. In this sense, we show how alternative states and regime shifts result from a very simple first principle, namely, nonlinear feedback mechanisms. We show that a model that has these feedbacks -- and not much more than that -- provides a more plausible description of observed data. In doing so we aim to highlight key properties that water reservoir share with other socio-environmental systems, which can contribute to fully-fledged models. 

In this study we focused on a specific reservoir, the Cantareira system in the metropolitan area of S\~ao Paulo.  Let us now elaborate on this specific case.  Signals of below the average rainfall were present since mid-2013, with corresponding decrease in water inflow. The conditional variance increased steadily in the very beginning of 2014, indicating an ongoing regime shift.  The average inflow rate in January 2014 was 15.7~$m^3/s$, whereas the outflow was kept at 34.3~$m^3/s$, a very unusual deficit for the rainy season in the region. Actually, the 30-day moving average of inflows and outflows presented deficits continually from May 2013 to February 2015, contrary to the recurrent pattern where the period from December-May always had greater inflows than outflows. 
The operators began decrementing releases only in March 2014, and at a pace that was not enough to bring back the system 
to the normal operating conditions. The obvious question is why the management policies could not avoid the collapse of the system.
Official documents available to the public \citep{ANA1} show that
the Brazilian Water Agency (\emph{ANA}) and the S\~ao Paulo State Water Department (\emph{DAEE}) 
agreed that operators (\emph{SABESP} and other water companies) could take up to 36~$m^3s{-1}$ of water 
from the Cantareira system. 
The agreement also included a reservoir rule curve that prescribed maximum allowed withdrawals according to
the stored volume of the system. These were not respected since January 2014.
Even if the limits were met, they were probably too loose. For instance, the outflow allowed for the rainy months of
November/December 2014 would be 27~$m^3s{-1}$ even if the operational capacity had been completely depleted. 
Since March 2014, \emph{ANA} and \emph{DAEE} abandoned the rule curve and adopted tighter limits. 
The maximum values have been negotiated periodically, and are currently 17.5~$m^3s{-1}$ (June-August 2015) 
and 13.5~$m^3s{-1}$ (September-November 2015). 
Nevertheless, the National Center for Surveillance of Natural Disasters (\url{http://www.cemaden.gov.br/})
forecasts that the Cantareira system will stay below its operational capacity at least until the beginning of 2016.

In summary, our results and our study case show that the management of reservoirs should take alternative regimes into account and avoid a transition to low-volume regimes. Failing to do so represents a prolongated burden, extending well beyond the period of anomalous rainfall, because outflow has to be kept as low as possible until a backwards transition occurs. Therefore managers should act as another feedback mechanism in the socio-environmental system that keeps it in the desired regime notwithstanding external forces like climate anomalies. In not doing so, managers of the Cantareira system acted like one more external force that pushed the reservoir to a catastrophic shift.

\section*{Data Sets}

Data for the time series of stored water volume in the main reservoir system of S\~ao Paulo State (Cantareira) and for the corresponding rainfall index were obtained from the  S\~ao Paulo Water Company (SABESP) database, which can de accessed at \url{http://www2.sabesp.com.br/mananciais/DivulgacaoSiteSabesp.aspx}.  Data for the inflow and outflow were obtained from \url{http://www2.sabesp.com.br/mananciais/Divulgacaopcj.aspx} and processed through Optical Character Recognition (OCR) software. Starting January 15th, 2015, a more convenient tool has been provided by SABESP at \url{http://site.sabesp.com.br/site/interna/Default.aspx?secaoId=553}. All data and codes used in the analyses are available at \url{https://github.com/cantareira/plos}.

\section*{Acknowledgments}
The authors thank FAPESP, CNPq and CAPES (Brazil) for partial support.

\bibliographystyle{authordate1}
\bibliography{bibliography}

\begin{thebibliography}{}

\bibitem[\protect\citename{{Ag\^encia Nacional de \'Aguas} \& {Departamento de
  \'Aguas e Energia El\'etrica}, }2015]{ANA1}
{Ag\^encia Nacional de \'Aguas}, \& {Departamento de \'Aguas e Energia
  El\'etrica}. 2015 (June).
\newblock {\em Dados de refer\^encia acerca da outorga do sistema Cantareira}.
\newblock
  \url{http://arquivos.ana.gov.br/institucional/sof/Renovacao_Outorga/DDR_Sistema_Cantareira
  - 12Jun15 - FINAL.pdf}.

\bibitem[\protect\citename{Beven, }2011]{beven2011}
Beven, Keith~J. 2011.
\newblock {\em Rainfall-runoff modelling: the primer}.
\newblock John Wiley \& Sons.

\bibitem[\protect\citename{Boettiger \& Hastings, }2012]{boettiger2012}
Boettiger, Carl, \& Hastings, Alan. 2012.
\newblock Quantifying limits to detection of early warning for critical
  transitions.
\newblock {\em Journal of The Royal Society Interface}, {\bf 9}(75),
  2527--2539.

\bibitem[\protect\citename{Boettiger {\em et~al.\ }\relax,
  }2013]{boettiger2013}
Boettiger, Carl, Ross, Noam, \& Hastings, Alan. 2013.
\newblock Early warning signals: the charted and uncharted territories.
\newblock {\em Theoretical ecology}, {\bf 6}(3), 255--264.

\bibitem[\protect\citename{Carpenter \& Brock, }2011]{carpenter2011}
Carpenter, SR, \& Brock, WA. 2011.
\newblock Early warnings of unknown nonlinear shifts: a nonparametric approach.
\newblock {\em Ecology}, {\bf 92}(12), 2196--2201.

\bibitem[\protect\citename{Carpenter, }2005]{carpenter2005}
Carpenter, Stephen~R. 2005.
\newblock Eutrophication of aquatic ecosystems: bistability and soil
  phosphorus.
\newblock {\em Proceedings of the National Academy of Sciences of the United
  States of America}, {\bf 102}(29), 10002--10005.

\bibitem[\protect\citename{Dakos, }2015]{dakosews}
Dakos, Vasilis. 2015.
\newblock {\em Early warning signals toolbox}.
\newblock \url{http://www.early-warning-signals.org}.
\newblock accessed on 29 Jan 2015.

\bibitem[\protect\citename{Dakos {\em et~al.\ }\relax, }2012]{dakos2012}
Dakos, Vasilis, Carpenter, Stephen~R, Brock, William~A, Ellison, Aaron~M,
  Guttal, Vishwesha, Ives, Anthony~R, Kefi, Sonia, Livina, Valerie, Seekell,
  David~A, van Nes, Egbert~H, {\em et~al.\ }\relax. 2012.
\newblock Methods for detecting early warnings of critical transitions in time
  series illustrated using simulated ecological data.
\newblock {\em PloS one}, {\bf 7}(7), e41010.

\bibitem[\protect\citename{Dakos {\em et~al.\ }\relax, }2015]{dakos2015}
Dakos, Vasilis, Carpenter, Stephen~R, van Nes, Egbert~H, \& Scheffer, Marten.
  2015.
\newblock Resilience indicators: prospects and limitations for early warnings
  of regime shifts.
\newblock {\em Philosophical Transactions of the Royal Society of London B:
  Biological Sciences}, {\bf 370}(1659), 20130263.

\bibitem[\protect\citename{Folke, }2006]{folke2006}
Folke, Carl. 2006.
\newblock Resilience: The emergence of a perspective for social--ecological
  systems analyses.
\newblock {\em Global environmental change}, {\bf 16}(3), 253--267.

\bibitem[\protect\citename{Gordon {\em et~al.\ }\relax, }2008]{gordon2008}
Gordon, Line~J, Peterson, Garry~D, \& Bennett, Elena~M. 2008.
\newblock Agricultural modifications of hydrological flows create ecological
  surprises.
\newblock {\em Trends in Ecology \& Evolution}, {\bf 23}(4), 211--219.

\bibitem[\protect\citename{King {\em et~al.\ }\relax, }2009]{pomp}
King, Aaron~A, Ionides, Edward~L, Bret{\'o}, CM, Ellner, Steve, Kendall, Bruce,
  Wearing, Helen, Ferrari, Matthew~J, Lavine, Michael, \& Reuman, Daniel~C.
  2009.
\newblock {\em pomp: Statistical inference for partially observed Markov
  processes}.
\newblock \url{http://pomp.r-forge.r-rproject.org}.

\bibitem[\protect\citename{Klausmeier, }1999]{klausmeier1999}
Klausmeier, Christopher~A. 1999.
\newblock Regular and irregular patterns in semiarid vegetation.
\newblock {\em Science}, {\bf 284}(5421), 1826--1828.

\bibitem[\protect\citename{Liu \& West, }2001]{liu2001}
Liu, Jane, \& West, Mike. 2001.
\newblock Combined parameter and state estimation in simulation-based
  filtering.
\newblock {\em Pages  197--223 of:} {\em Sequential Monte Carlo methods in
  practice}.
\newblock Springer.

\bibitem[\protect\citename{Meron {\em et~al.\ }\relax, }2004]{meron2004}
Meron, Ehud, Gilad, Erez, von Hardenberg, Jost, Shachak, Moshe, \& Zarmi, Yair.
  2004.
\newblock Vegetation patterns along a rainfall gradient.
\newblock {\em Chaos, Solitons \& Fractals}, {\bf 19}(2), 367--376.

\bibitem[\protect\citename{{\"O}sterblom {\em et~al.\ }\relax,
  }2013]{osterblom2013}
{\"O}sterblom, Henrik, Merrie, Andrew, Metian, Marc, Boonstra, Wiebren~J,
  Blenckner, Thorsten, Watson, James~R, Rykaczewski, Ryan~R, Ota, Yoshitaka,
  Sarmiento, Jorge~L, Christensen, Villy, {\em et~al.\ }\relax. 2013.
\newblock Modeling Social-Ecological Scenarios in Marine Systems.
\newblock {\em BioScience}, {\bf 63}(9), 735--744.

\bibitem[\protect\citename{Scheffer \& Carpenter, }2003]{scheffer2003}
Scheffer, Marten, \& Carpenter, Stephen~R. 2003.
\newblock Catastrophic regime shifts in ecosystems: linking theory to
  observation.
\newblock {\em Trends in ecology \& evolution}, {\bf 18}(12), 648--656.

\bibitem[\protect\citename{Scheffer \& van Nes, }2007]{scheffer2007}
Scheffer, Marten, \& van Nes, Egbert~H. 2007.
\newblock Shallow lakes theory revisited: various alternative regimes driven by
  climate, nutrients, depth and lake size.
\newblock {\em Hydrobiologia}, {\bf 584}(1), 455--466.

\bibitem[\protect\citename{Scheffer {\em et~al.\ }\relax, }2001]{scheffer2001}
Scheffer, Marten, Carpenter, Steve, Foley, Jonathan~A, Folke, Carl, \& Walker,
  Brian. 2001.
\newblock Catastrophic shifts in ecosystems.
\newblock {\em Nature}, {\bf 413}(6856), 591--596.

\bibitem[\protect\citename{Young {\em et~al.\ }\relax, }2004]{young2004}
Young, Peter~C, Chotai, Arun, \& Beven, Keith~J. 2004.
\newblock Data-based mechanistic modelling and the simplification of
  environmental systems.
\newblock {\em Pages  371--388 of:} Wainwright, John, \& Mulligan, Mark (eds),
  {\em Environmental Modelling: Finding Simplicity in Complexity}.
\newblock Chichester, UK: Wiley.

\end{thebibliography}

\end{document}